\documentclass{jfm}
\usepackage{graphicx}
\usepackage{amsmath}
\usepackage{float}
\usepackage[figuresleft]{rotating}
\usepackage{lscape}
\usepackage{caption}
\usepackage{subcaption}
\usepackage{epstopdf, epsfig}

\usepackage{xcolor}
\usepackage{color}
\newcommand{\B} [1] {{\bf {#1}}}

\definecolor{brick}{rgb}{0.75,0,0}
\definecolor{rltgreen}{rgb}{0,0.5,0}

\newlength{\wdth}

\shortauthor{A. Jouin, S. Cherubini, J.-C. Robinet}

\title{Collective secondary instabilities: an application to three-dimensional boundary-layer flow}

\author{A. Jouin \aff{1} \aff{2},
  S. Cherubini \aff{2}
 \and J.-C. Robinet\aff{1}}

\affiliation{
\aff{1} DynFluid, Ecole Nationale Supérieure des Arts et Métiers, 75013 Paris, France
\aff{2} Dipartimento di Meccanica, Matematica e Management, Politecnico di Bari, 70126 Bari, Italy
}

\begin{document}

\maketitle

\begin{abstract}
In some linearly unstable flows, secondary instability is found to have a much larger wavelength than that of the primary unstable modes, so that it cannot be recovered with a classical Floquet analysis.  In this work, we apply a new formulation for capturing secondary instabilities coupling multiple length scales of the primary mode. This formulation, based on two-dimensional stability analysis coupled with a Bloch waves formalism originally described in \citet{schmid_periodic_2017}, allows to consider high-dimensional systems resulting from several repetitions of a periodic unit, by solving an eigenproblem of much smaller size. Collective instabilities coupling multiple periodic units can be thus retrieved. The method is applied on the secondary stability of a swept boundary-layer flow subject to stationary cross-flow vortices, and compared with Floquet analysis. Two multi-modal instabilities are recovered: for streamwise wavenumber $\alpha_v$ close to zero, approximately twelve sub-units are involved in large-wavelength oscillations; whereas a staggered pattern, characteristic of subharmonic instabilities, is observed for $\alpha_v = 0.087$.
\end{abstract}

\section{Introduction}


Transition to turbulence is a crucial problem  of great technical interest in  fluid mechanics. A successful approach for studying it consists in dividing the process into several stages. An initial stage, denoted as primary instability, can be investigated through linear stability theory, either modal or non-modal \citep{schmid_stability_2001}. These analyses yield structures representative of the primary instability mechanisms such as, for example, Tollmien-Schlichting (TS) waves for channel or boundary-layer flows \citep{Tollmien1930}, cross-flow vortices for swept flows \citep{mack_boundary-layer_1984}, gravity waves in stratified flows \citep{chandrasekhar_1955} or streaks in the case of non-modal mechanisms \citep{reddy_energy_1993}.
The previous list is far from being exhaustive as the relevant structures strongly depend on the flow configuration and the case considered. 
\\
These flow structures are likely to further destabilise, resulting in the second stage of the transition process. Investigation of this second stage is possible through secondary stability theory, established through the seminal works of \citet{orszag_patera_1983} and \citet{herbert_secondary_1985} on two-dimensional TS waves in wall-bounded flows. Precisely, in the simplest cases, secondary stability theory is based on a Floquet analysis of a secondary base flow constructed as the superposition of an initial laminar flow and a primary disturbance previously identified. Notably, in channel and boundary layer flows, the exponential growth of small three-dimensional disturbances, observed experimentally and in numerical simulations, was explained as a secondary instability of two-dimensional TS waves \citep{orszag_patera_1983,herbert_secondary_1985}. Following these seminal works, the theory was applied to numerous other configurations such as, for example, streaks in boundary-layer \citep{liu_floquet_2008} or cross-flow vortices in swept flows \citep{fischer_primary_1991, bonfigli_secondary_2007}, leading to breakthroughs in the identification of the underlying mechanisms of transition to turbulence.
\\
However, Floquet analysis may not be sufficient in cases where the secondary instability has a much larger wavelength than the initial wavelength of the primary disturbance. Such a peculiar behaviour can be observed in several flow configurations, usually coupling multiple length scales, such as the flow over superhydrophobic surfaces with moving interfaces \citep{picella_influence_2020}, large arrays of vortices \citep{tabeling_cardoso_perrin_1990} or vortex pairings in mixing layers \citep{ho_huang_1982}.
Unfortunately, the classical Floquet analysis successfully recovers modal instabilities that are periodic over individual primary instability units (i.e., one superhydrophobic roughness, one vortex pair etc), but does not allow to recover modes that are periodic only over an array of $n$ such units, while not on the single sub-unit.
\\
To overcome this limitation, we propose in this article to apply a new formulation for secondary stability analysis based on the association of a local two-dimensional stability analysis and a Bloch waves formalism. 
This  formulation is based on the mathematical framework proposed by \citet{schmid_periodic_2017}  for the primary stability analysis of fluid systems consisting of a periodic array of $n$ identical units. This method allows to compute multi-modal fluid instabilities, i.e., modal solutions that are periodic over the array of $n$ units, despite not being periodic over individual units, without solving the whole eigenvalue problem.
\\
We present here an application of this framework to secondary stability theory of fluid flows, which provides a great benefit on its computational cost. In this context, the periodicity of the base flow, which is necessary for the application of the method, results from the periodicity of the primary disturbance, and not from the spatial recurrence of a geometric feature as for the examples provided in \citet{schmid_periodic_2017}. Using this formalism, secondary stability analyses can be carried out on very large systems, allowing to capture resonances or complex multi-modal behaviour spanning several wavelengths of the primary disturbance. This type of global behaviour involving multiple units, will be referred as \textit{collective}, to prevent confusion with the commonly accepted terminology of global modes \citep{theofilis_2011}. The term \textit{collective instability} has been used in \citet{ho_huang_1982}, referring to the interaction of a large number of vortices in subharmonic transition in mixing layers.
\\
As a proof of concept, the proposed method is applied on the secondary stability of a swept boundary layer subject to stationary cross-flow modes. Several reasons motivate this choice: first, a direct comparison can be made with results from the Floquet analysis of \citet{fischer_primary_1991}. Second, in a DNS study from \citet{hogberg_secondary_1998}, some collective behaviour of the secondary instability appeared, but it was not further investigated. Finally, to the authors' knowledge, despite the extensive literature on the secondary instabilities in swept boundary layers \citep{fischer_primary_1991,koch_secondary,wassermann_mechanisms_2002,bonfigli_secondary_2007,serpieri_three-dimensional_2016}, no study has considered yet the simultaneous secondary instability of a large number of cross-flow vortices. 
\\
The outline of the remainder of the paper is as follows. The mathematical framework described in \citet{schmid_periodic_2017} will be adapted for the study of secondary instabilities in \S\ref{sec:THEORY}. The results and discussion on the secondary stability of the swept boundary layer are contained in \S \ref{sec:RESULTS}. The main findings are summarised in \S \ref{sec:CONCLUSIONS}.

\section{Mathematical framework}\label{sec:THEORY}

Let us consider a general, time-evolving system of the form:
\begin{equation}
    \frac{\partial \mathbf{Q}}{ \partial t} = \mathbf{F}(\B{x},t;\mathbf{Q})
    \label{EqSys}
\end{equation}

where $\mathbf{Q}$ is the state vector, $\mathbf{F}$ a non-linear operator and $\B{x}=(x,y,z)$ is the spatial coordinates vector. The flow is decomposed into a base flow $\mathbf{Q}_0\left(y\right)$, assumed to be locally parallel, and a small primary disturbance $\mathbf{q}_1(\B{x},t)$, such that $\mathbf{Q}(\B{x},t) = \mathbf{Q}_0(y) + \mathbf{q}_1(\B{x},t)$. Substituting the previous decomposition into Eq.(\ref{EqSys}) and neglecting nonlinear terms results in the following system for the perturbation:
\begin{equation}
    \frac{\partial \mathbf{q}_1}{\partial t} = \mathbf{A} \mathbf{q}_1
\end{equation}
with $\mathbf{A}$ the Jacobian of the system. Assuming normal modes of the form $\mathbf{q}_1(\mathbf{x},t) = \Tilde{\mathbf{q}}_1(y)\exp{[i(\alpha x + \beta z  - \omega t)]}$, modal and non modal stability analysis can be performed, identifying specific waves of interest, with wavenumbers $(\alpha,\beta)$ and frequency $\omega$.

From there, following \citet{herbert_secondary_1985}, the secondary base flow $\mathbf{Q}_1$ is defined as the superposition of the primary disturbance of interest $\mathbf{q}_1$, with amplitude $A$, and the primary base flow $\mathbf{Q}_0$, such that $\mathbf{Q}_1(\mathbf{x},t) = \mathbf{Q}_0(y) + A\mathbf{q}_1(\mathbf{x},t)$. We also introduce a new Galilean coordinate system $(x_v,y,z_v)$ normal to the wave vector $ \mathbf{k} = (\alpha, \beta)^T$ of the primary disturbance moving with phase speed $c=\omega_i/||\mathbf{k}||$ in the $z_v$-direction. The passage from one coordinate system to the other can be performed through the following Squire transform: $x_v(t) = \beta/||\mathbf{k}||x - \alpha/||\mathbf{k}||z$ and $z_v(t) = \alpha/||\mathbf{k}||x + \beta/||\mathbf{k}||z$ - ct. In this new frame, the secondary base flow $\mathbf{Q}_1(y,z_v)$ is stationary, streamwise independent and $2\pi/||\mathbf{k}||$-periodic in the spanwise direction $z_v$. In the wave-oriented reference frame, the flow is decomposed into the secondary base flow $\mathbf{Q}_1(y,z_v)$ and a small secondary perturbation $\mathbf{q}_2(x_v,y,z_v,t)$ such as $\mathbf{Q}(x_v,y,z_v,t) = \mathbf{Q}_1(y,z_v) + \mathbf{q}_2(x_v,y,z_v,t)$. This formulation requires the shape assumption to be valid. Potentially, the base flow for secondary stability analysis can be retrieved by other means, such as parabolised stability equations or direct extraction from a numerical simulation. The key point is to guarantee the periodicity of the secondary base flow on the sub-units. Modal secondary perturbations are assumed: $\mathbf{q}_2(x_v,y,z_v,t) = \Tilde{\mathbf{q}}_2(y,z_v)\exp{(i(\alpha_vx_v -\sigma t))}$ with $\alpha_v$ and $\sigma$ being respectively the streamwise wavenumber and the circular frequency. 

In the classical theory of secondary stability, Floquet analysis would be applied on a periodic domain consisting on only one wavelength $2\pi/||\mathbf{k}||$ of the base flow  $\mathbf{Q}_1$. This method recovers instabilities that are periodic over individual units, but does not allow to compute modes that are periodic only over the array of $n$ units (and not on the single sub-unit). Instead, we consider a fluid system composed of the repetition in the spanwise direction $z_v$ of the secondary base flow $\mathbf{Q}_1$ over $n$ sub-units of length $2\pi/||\mathbf{k}||$.  The Navier-Stokes equations are then linearised around the $2\pi/||\mathbf{k}||$-periodic base flow $\mathbf{Q}_1(y,z_v)$. The ansatz for the secondary disturbance is introduced, yielding a two-dimensional local stability problem with $2\pi/||\mathbf{k}||$-periodic coefficients. The equations for the stability problem linearised around a two-dimensional stationary base flow $\mathbf{Q}_1(y,z_v) = [U(y,z_v),V(y,z_v),W(y,z_v)]^T$ can be found in \citet{loiseau_dynamics_2014}.

Performing a classical linear stability analysis of a system of this size yields a prohibitive computational cost. However, this task can be tackled at an affordable cost using the framework described by \citet{schmid_periodic_2017}. For the sake of completeness, the method will be quickly described in the following. The first step consists in reordering the system, which is partitioned into $n$ smaller systems, each one corresponding to a sub-unit. Mathematically, the disturbance equations can be recast under the following form:  

\begin{equation}
     \frac{\partial}{\partial t}
     \begin{pmatrix}
           \mathbf{q}_2^0 \\
           \mathbf{q}_2^1 \\
           \vdots \\
           \mathbf{q}_2^{n-1}
         \end{pmatrix}
         =
        \underbrace{\begin{pmatrix}
           \mathbf{A}_0 & \mathbf{A}_1 & \hdots & \mathbf{A}_{n-1}      \\
           \mathbf{A}_{n-1} & \mathbf{A}_0 & \hdots & \mathbf{A}_{n-2}  \\
           \vdots & \vdots & & \vdots \\
           \mathbf{A}_1 & \mathbf{A}_2 & \hdots & \mathbf{A}_0
         \end{pmatrix}}_{\mathbf{A'}}
        \underbrace{\begin{pmatrix}
           \mathbf{q}_2^0 \\
           \mathbf{q}_2^1 \\
           \vdots \\
           \mathbf{q}_2^{n-1}
         \end{pmatrix}}_{\mathbf{q}_2}
\end{equation}

with $\mathbf{A'}$ the Jacobian associated to the full secondary stability problem, which is composed of the matrices $\mathbf{A}_0$ and $\mathbf{A}_j$ (for $j=1, ...,n-1)$ describing respectively the dynamics in a sub-unit and the coupling interactions between sub-units. The secondary disturbance state vector in the $j^{\text{th}}$ sub-unit is denoted as $\mathbf{q}_2^j$. The Jacobian matrix $\mathbf{A'}$ is block-circulant due to the specific $n$-periodic nature of the system and can become block-diagonal through the similarity transformation:

\begin{equation}
    \mathbf{P}^H \mathbf{A'} \mathbf{P} = \text{diag}(\Hat{\mathbf{A}}_0, \Hat{\mathbf{A}}_1, ... , \Hat{\mathbf{A}}_{n-1}  ) \equiv \hat{\mathbf{A}}
\end{equation}

The transfer matrix $\mathbf{P}$ can be found analytically as:
\begin{equation}
    \mathbf{P} = \mathbf{J} \otimes \mathbf{I}
\end{equation}

with $\mathbf{J}$ a matrix such as $\mathbf{J}_{j+1,k+1} = \rho_j^k/\sqrt{n}$ for $j,k=1, ..., n-1$ and $\rho_j = \exp(2i\pi j/n)$ the $n^{\text{th}}$ roots of unity. The symbol $\otimes$ denotes the usual Kronecker product and $\mathbf{I}$ the identity matrix. With this transformation, the linear stability problem has been reduced to the study of $n$ smaller sub-systems characterised by the matrices $\Hat{\mathbf{A}}_j$. Hence, the full spectrum of the matrix $\mathbf{A}$ can be found  merging the $n$ spectra of $\hat{\mathbf{A}}_j$ for $j=1,...,n-1$. Similarly, provided $\mathbf{v}_j$ is an eigenvector of $\hat{\mathbf{A}}_j$, the eigenfunctions of the full system can be retrieved and take the form $[\mathbf{v}_j, \rho_j \mathbf{v}_j, \rho_j^2 \mathbf{v}_j, ..., \rho_j^{n-1} \mathbf{v}_j]^T$ for $j=1,...,n-1$. Physically, $\rho_j$ acts as a phase shift between the different sub-units: the farther it is from $0$ (or $2\pi$), the more desynchronised the collective mode is.

In the case of nearest-neighbour coupling, rather common in many applications, the Jacobian $\mathbf{A'}$ reduces to a block-tridiagonal matrix. Only a three-unit system $\mathbf{A}_{0},\mathbf{A}_{1},\mathbf{A}_{2}$ needs to be discretized and processed, significantly reducing the complexity and computational cost of the method. Both temporal and spatial approaches for the stability problem can be considered. 
Finally, one can notice that non-modal and resolvent analyses can also be carried out within this framework. These have not been described as the present article focuses on modal instabilites. For the full discussion, the reader is referred to \citet{schmid_periodic_2017}. 

\section{Results}\label{sec:RESULTS}

To illustrate the previous method, the secondary instability of stationary cross-flow vortices in a swept-boundary flow is investigated. This analysis extends and completes the work of \citet{fischer_primary_1991}. All the stability results hereafter are obtained using a temporal approach ($\omega, \sigma \in \mathbb{C}$). The secondary stability problem is tackled considering the nearest-neighbour coupling assumption. For both primary and secondary analyses, the stability problem is discretised with a spectral collocation method \citep{trefethen_spectral_2000}. Wall-normal and spanwise directions are respectively discretised with a $64$-points Fourier grid and a $64$-points Chebyshev grid. The generalised eigenvalue problem is solved using a Krylov-Schur algorithm of the SciPy Python module coupled with a shift-and-inverse technique. 

\subsection{Primary stability analysis}\label{sec:PRIMARY}

\begin{figure}
    \centering
    \includegraphics[width=0.42\textwidth]{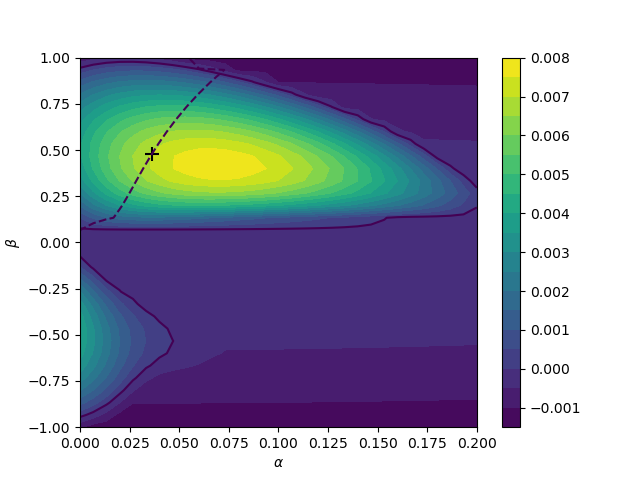}
    \includegraphics[width=0.42\textwidth]{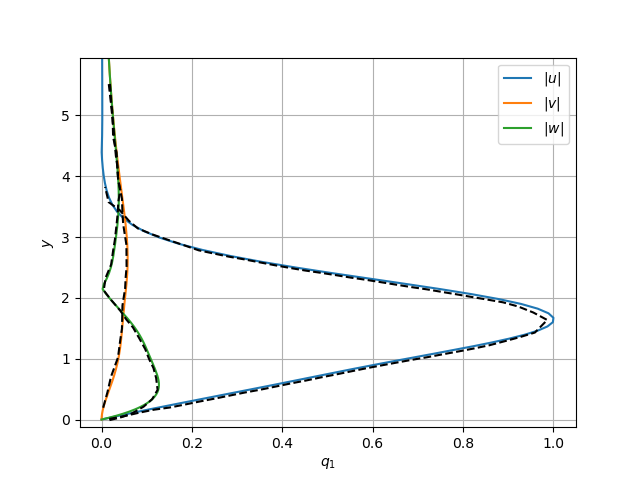}
    \caption{Left: Neutral curve of the swept boundary layer for $Re_{\delta} = 826$. The dashed line corresponds to stationary disturbances i.e. $\omega_i = 0$. Right: Absolute value of the eigenfunctions for $Re_{\delta} = 826$, $\alpha = 0.0361$ and $\beta = 0.4774$ ($\Psi = \arctan \left(\beta/\alpha\right)= 89.9^o$). The dashed lines are extracted from \citet{fischer_primary_1991}. }
    \label{fig:Primary}
\end{figure}

\begin{figure}
    \centering
    \includegraphics[width=0.42\textwidth]{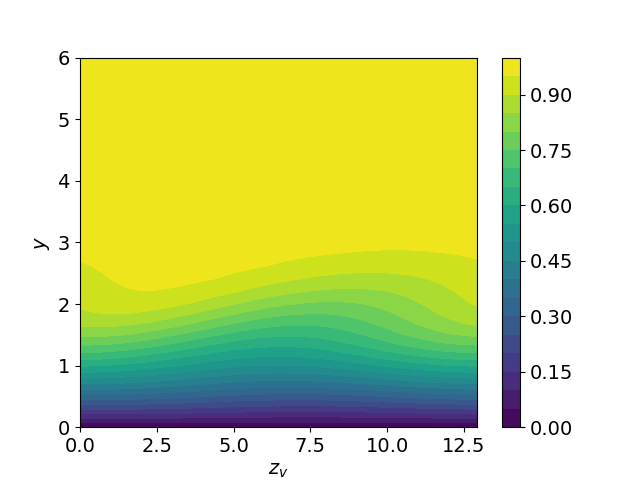}
    \includegraphics[width=0.42\textwidth]{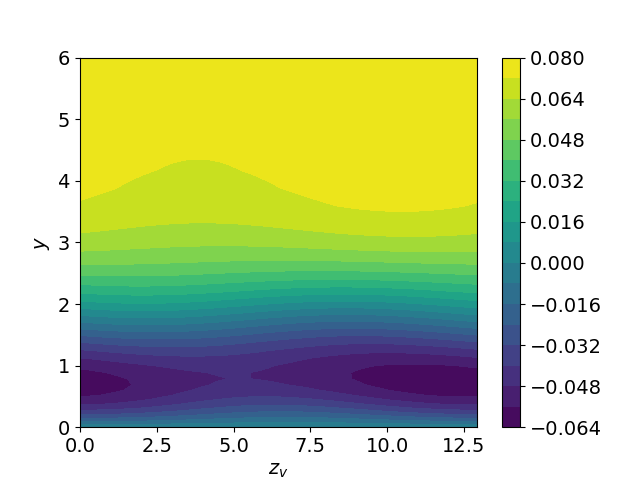}
    \caption{Secondary base flow $U_1(y,z_v)$ (left) and $W_1(y,z_v)$ (right) in the wave-oriented reference frame, for one sub-unit. The wall-normal component $V_1$ is non-zero but is one order of magnitude smaller than $W_1$, thus it is not shown.}
    \label{fig:BF}
\end{figure}

In the following, we consider the incompressible flow over an infinite swept flat plate with an imposed negative pressure gradient (i.e. decreasing pressure in the chordwise direction). The laminar base flow $\mathbf{Q}_0=[U_0(y),0,W_0(y)]^T$ is modelled with a Falkner-Skan-Cook profile \citep{FS1931,cooke1950}. Precisely, introducing $f$ and $g$ such that $U_0(y) = f'(y)\cos{\theta}$ and $W_0(y) = g(y)\sin{\theta}$, we have:
\begin{align}
    (2-\beta_H)f'''&+ ff'' + \beta_H\left[ 1 - \left(f'\right)^2 \right] = 0 \\
    (2-\beta_H)g'' &+ fg' = 0,
\end{align}

where the Hartree dimensionless pressure-gradient parameter ($\beta_H$), the local sweep angle and the local Reynolds number are set respectively to $\beta_H = 0.630$, $\theta = 46.9^{\circ}$ and $Re_{\delta} = 826$. These parameters are taken directly from \citet{fischer_primary_1991}, and set to fit qualitatively experimental results from \citet{muller_1990}. Assuming wave-like solutions for the disturbances, modal stability analysis is obtained for this configuration. Stationary disturbances maximising the temporal growth rate with respect to the streamwise and spanwise wavenumbers $\alpha$ and $\beta$ are sought. Figure \ref{fig:Primary} exhibits the neutral curve of the primary instability. Since Squire's theorem is not valid for three-dimensional flows \citep{pralits_stability_2017}, the $\beta<0$ plane was also investigated for the sake of completeness. The maximum growth rate for stationary disturbances is reached for $(\alpha,\beta) = (0.0361, 0.4774)$. The eigenfunctions of this mode are displayed in figure \ref{fig:Primary} (right). A second instability region appears as well, albeit with lower growth rates and no stationary disturbances. The agreement with the results found in \citet{fischer_primary_1991} is excellent.

\subsection{Secondary stability analysis}\label{sec:SECONDARY}

Secondary stability of the flow is investigated using the framework described in \S \ref{sec:THEORY}. The secondary base blow is defined as $\mathbf{Q}_1(y,z_v) = \mathbf{Q}_0(y) + A\mathbf{q}_1(y,z_v)$ with $\mathbf{q}_1$ the disturbance with wavevector $\mathbf{k} = (0.0361,0.4774)^T$ in the original reference frame. The amplitude $A$ is set as $A = 0.0789$ \citep{fischer_primary_1991}. The resulting base flow on one sub-unit and in the wave-oriented reference frame is shown in figure \ref{fig:BF}. Qualitative validity of the shape assumption in this particular case is shown through comparison with experiments in \citet{fischer_primary_1991}.

\begin{figure}
\begin{tabular}{cc}
    \begin{minipage}{0.6\textwidth} \includegraphics[width=1\textwidth]{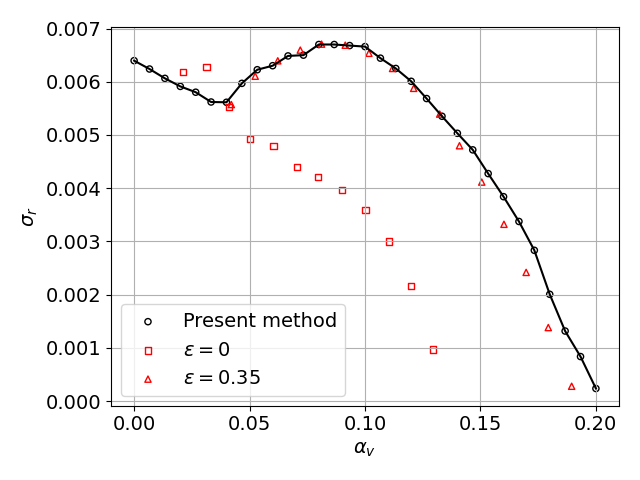} \end{minipage}& \begin{minipage}{0.4\textwidth} \includegraphics[width=0.73\textwidth]{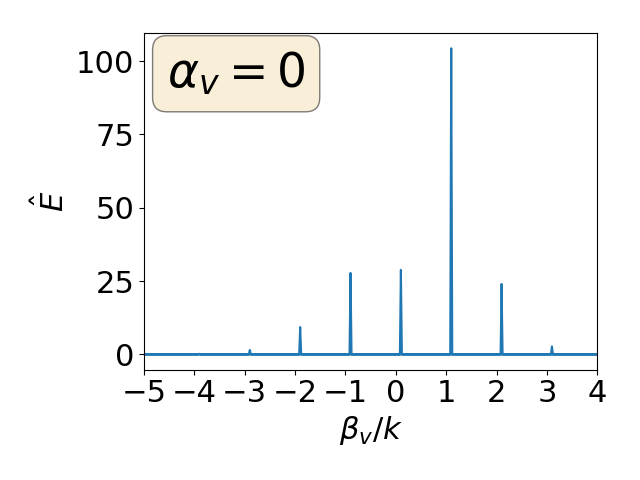} \\ \includegraphics[width=0.73\textwidth]{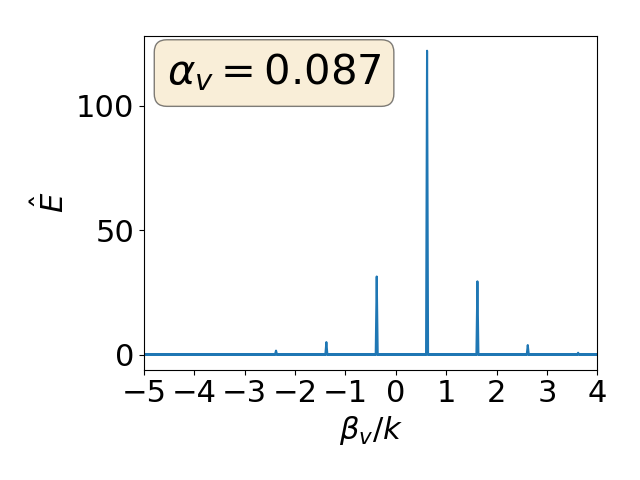} \end{minipage}
    \end{tabular}
    \caption{Left: evolution of the secondary growth rate $\sigma_r$ (black circles) as a function of the streamwise wavenumber $\alpha_v$. A comparison is made with figure 8 of \citet{fischer_primary_1991} (red symbols). $\epsilon$ corresponds to the detuning factor of Fischer's Floquet analysis. Right: Spatial Fourier spectra of the energy of the most unstable mode for $\alpha_v = 0$ (top) and $\alpha_v = 0.087$ (bottom).}
    \label{fig:GrowthRate}
\end{figure}

Secondary growth rate for the full system as a function of the streamwise wavelength is displayed in figure \ref{fig:GrowthRate}. Two modes appear to compete: the maximum growth rate $\sigma_r = 0.0068$ is reached for $\alpha_v = 0.087$ while streamwise independent perturbations ($\alpha_v=0$) also display strong amplification with $\sigma_r = 0.0065$. A direct comparison can be made with figure 8 from \citet{fischer_primary_1991}, where Floquet analysis (on one sub-unit) is conducted for both harmonic modes and resonant modes with a detuning factor $\epsilon=0.35$, where $\epsilon = \gamma_i/||\mathbf{k}||$ and $\gamma_i$ is the imaginary part of the Floquet exponent $\gamma$. Maximum growth rates are obtained for $\alpha_v=0.03$ and $\alpha_v=0.08$, respectively. The agreement with the $\epsilon=0.35$ curve is good although some discrepancies are observed for the growth rates of harmonic modes. The Floquet analysis seems to overestimate the secondary growth rate in the range where small wavenumber modes are predominant. Also, the frequency of the most unstable mode for $\alpha_v=0.087$ is found to be $f=154$ Hz in relative good agreement with the frequency $f=145$ Hz found by Fischer. On the contrary, the unstable mode found for $\alpha_v = 0$ have a much lower frequency than the one obtained by \cite{fischer_primary_1991}, i.e., $f=4$ Hz instead of $f=74$ Hz.

The nature of the instabilities obtained in the new framework can be inferred inspecting the spatial Fourier energy spectra of the most unstable modes. These Fourier spectra, realised for the cases with $\alpha_v=0$ and $\alpha_v=0.087$, are shown in the right part of figure \ref{fig:GrowthRate}. The detuning factor can be identified, from the spectra, as $\epsilon = 1 - \beta_v^0/k$ with $\beta_v^0$ the wavenumber of the Fourier fundamental mode. Thus, the maximum at $\alpha_v=0.087$ is associated to a detuned mode with $\epsilon=0.35$, in agreement with \citet{fischer_primary_1991}. For $\alpha_v = 0$, all the frequencies are almost multiples of $k$, indicating a quasi-fundamental ($\epsilon=0.08$) nature of the instability. It is also worth noticing the larger number of modes required to accurately describe the instability. The secondary perturbation is truly multi-modal with important coupling effects between sub-units. In this context, accurate results may be harder to retrieve through a Floquet analysis, likely causing the discrepancies observed for the $\alpha_v=0$ modes.

\begin{figure}
    \centering
    \includegraphics[width=0.45\textwidth]{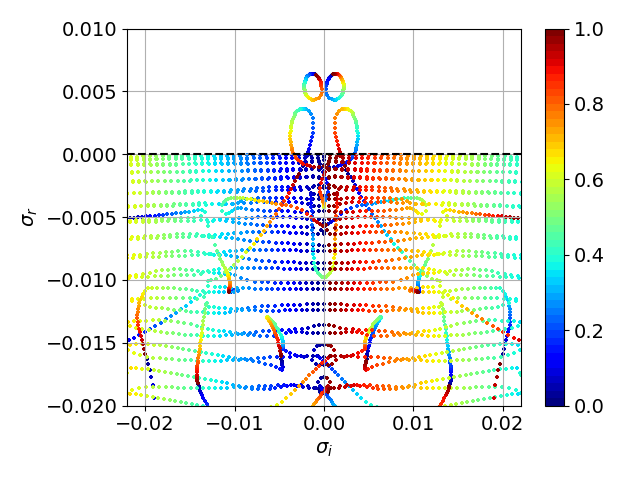}
    \includegraphics[width=0.45\textwidth]{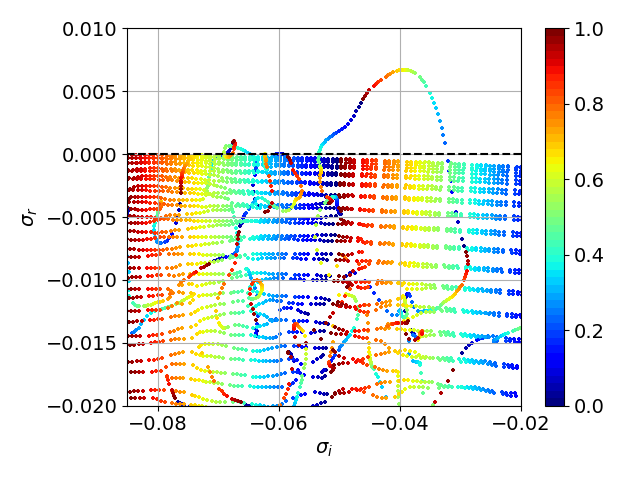}
    \caption{Full spectra of the secondary stability problem for $Re_{\delta} = 826$, $A = 0.0789$, $n=50$ and for $\alpha_v = 0$ (left) and $\alpha_v = 0.087$ (right). The full spectrum is constructed merging the $n$ spectra of the sub-systems. The eigenvalues have been coloured by the argument of their respective root of unity: $z \in \left[0,1\right]$ such as $\rho = \exp{(2i\pi z)}$. The dashed line corresponds to the $\sigma_r=0$ line.}
    \label{fig:Spectra}
\end{figure}

The spectra for the system composed of $n=50$ sub-units and for $\alpha_v = 0$ and $\alpha_v =  0.087$ are displayed in figure \ref{fig:Spectra}. A high number of sub-units corresponds to a (very) large system and allows a wide range of admissible spanwise wavenumbers. This explains the appearance of many branches within the spectra. The higher the number of sub-units, the more continuous these branches will be. Quasi-horizontal branches represent convective Squire modes and do not play a role in the asymptotic stability as they are always stable. The eigenvalues are coloured with the argument of their corresponding root-of-unity $\rho_j$. Brighter colours indicate important phase shift between sub-units and, thus, strong desynchronisation of the perturbation. For $\alpha_v = 0$, as expected, the spectrum is symmetric about the $\sigma_i = 0$ axis. Two unstable branches are found: the most unstable one loops on itself. The second remains open ended. The maximum growth rate is reached for $\rho_5=0.81+0.59j$, corresponding to a phase shift $\theta=36^o$ between the sub-units and indicating limited desynchronisation. The spectrum for $\alpha_v = 0.087$ is quite different. Two unstable branches are also observed. The first is marginally stable, and reaches a maximum growth rate for synchronised modes. The second is much more unstable and reaches a maximum for $\rho_{31} = -0.73-0.68j$, equivalent to a phase shift $\theta= -137^o$ and causing strong desynchronisation.

\begin{figure}
    \centering
    \begin{subfigure}[b]{0.35\textwidth}
         \centering
         \includegraphics[width=0.8\textwidth]{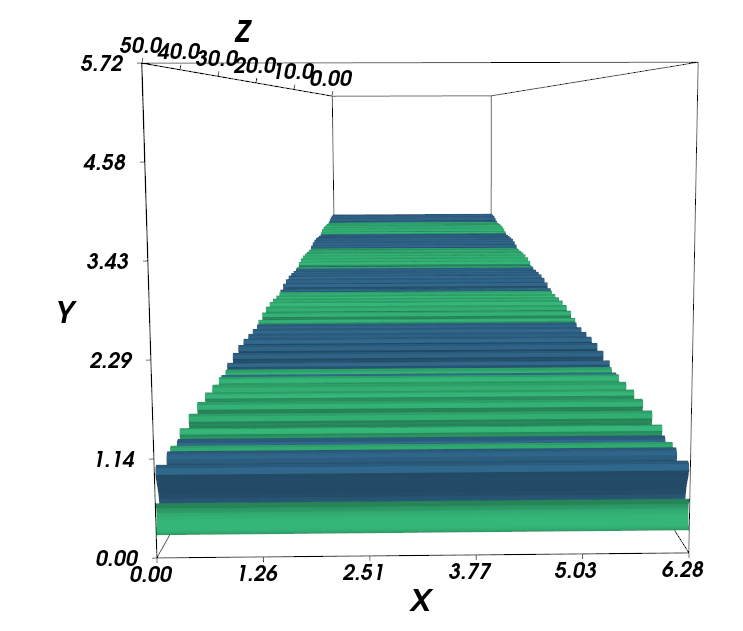}
         \caption{$\alpha_v=0$, $n=50$}
         \label{fig:y equals x}
    \end{subfigure}
    \begin{subfigure}[b]{0.45\textwidth}
         \centering
         \includegraphics[width=0.8\textwidth]{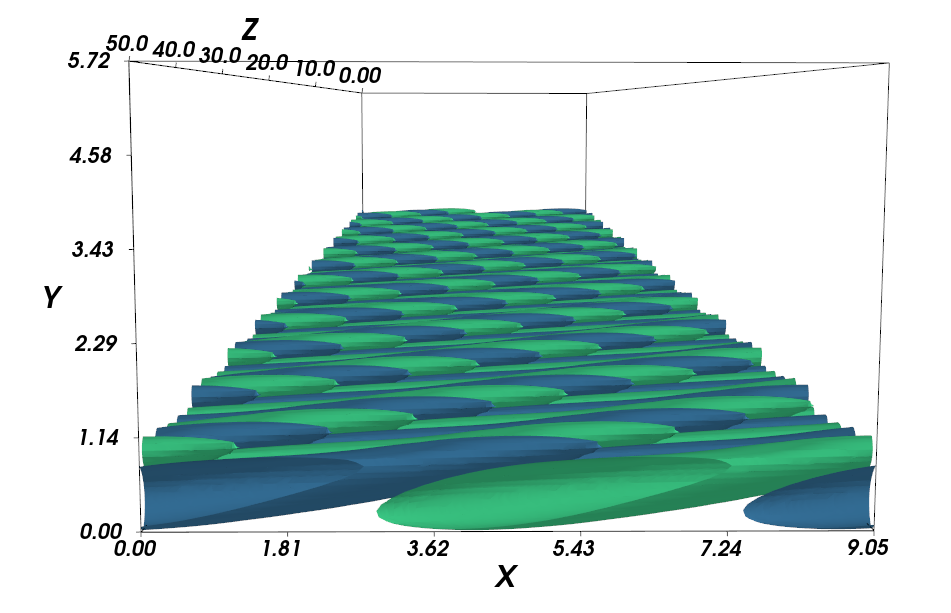}
         \caption{$\alpha_v=0.087$, $n=50$}
         \label{fig:y equals x}
    \end{subfigure}
    \begin{subfigure}[b]{0.95\textwidth}
         \centering
         \includegraphics[width=0.8\textwidth]{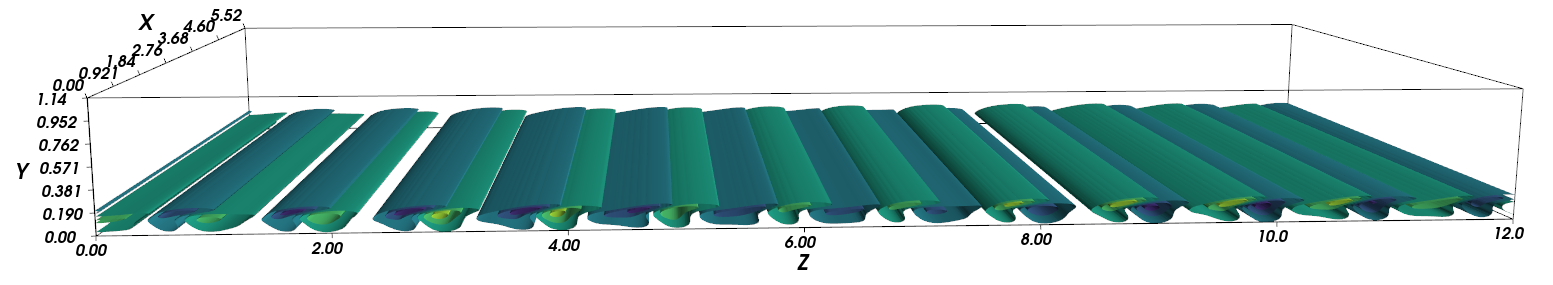}
         \caption{$\alpha_v=0$, $n=12$}
         \label{fig:y equals x}
    \end{subfigure}
    \begin{subfigure}[b]{0.95\textwidth}
         \centering
         \includegraphics[width=0.8\textwidth]{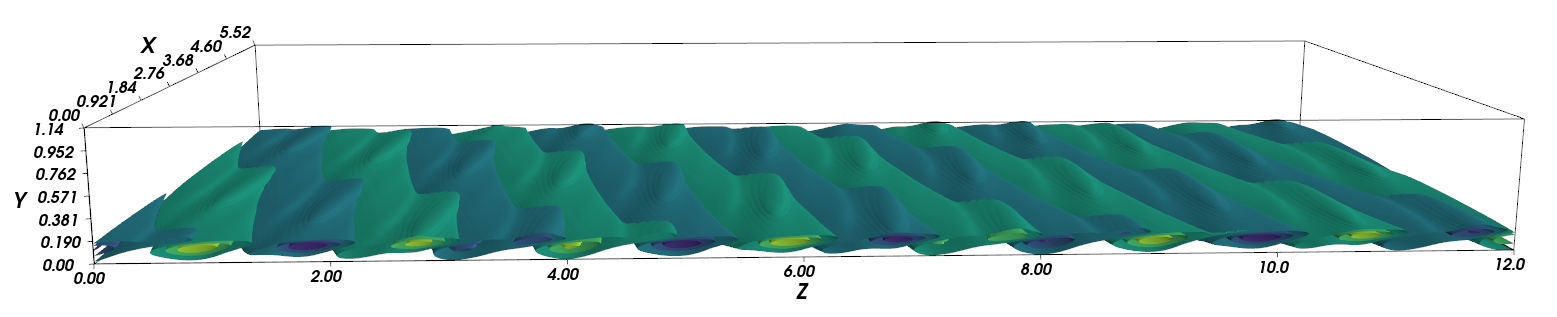}
         \caption{$\alpha_v=0.087$, $n=12$}
         \label{fig:y equals x}
    \end{subfigure}
    \caption{Three dimensional views of the streamwise velocity perturbation of the most unstable mode for different streamwise wavenumber $\alpha_v$ and number of visualised sub-units $n$. Notice the large wavelength instability for $\alpha_v=0$.}
    \label{fig:Mode3D}
\end{figure}

\begin{figure}
    \centering
    \includegraphics[width=0.49\textwidth]{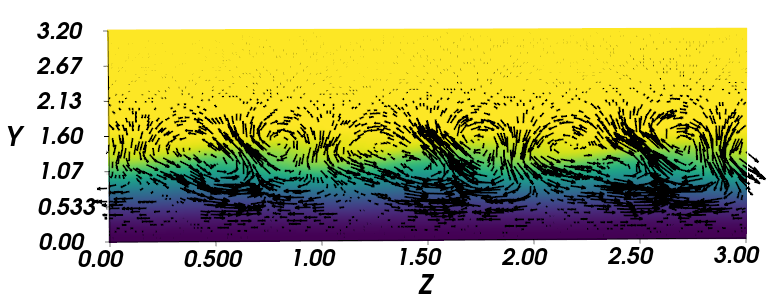}
    \includegraphics[width=0.49\textwidth]{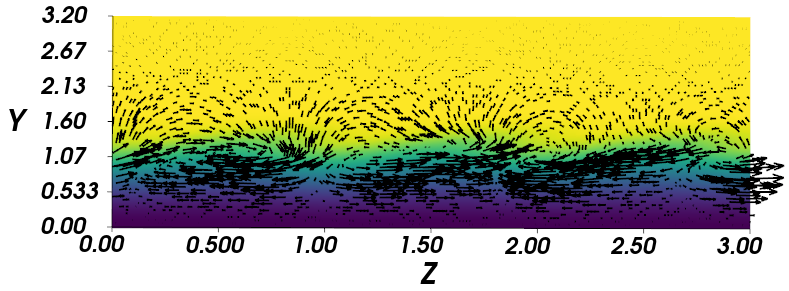}
    \caption{Vectors: cross-flow velocity in the plane $(y,z_v)$ for a group of three sub-units, with $\alpha_v = 0$ (left) and $\alpha_v=0.087$ (right). The contour plot represents the streamwise component of the secondary base flow. The flow in the $z$ direction goes from left to right.}
    \label{fig:CF}
\end{figure}

Ultimately, the most unstable modes are reconstructed for $\alpha_v = 0$ and $\alpha_v =  0.087$. Figure \ref{fig:Mode3D} displays the streamwise component of the disturbance velocity in the full domain while figure \ref{fig:CF} exhibits the cross-flow velocities for a group of three sub-units. The streamwise velocity disturbance is one order of magnitude higher than the cross-flow components. The position of the maxima of the streamwise disturbance slightly varies across the sub-units, although it is approximately located on the upwelling of the wave pattern of the secondary base flow. This is quite reminiscent of the \textit{type-I} modes of \citet{bonfigli_secondary_2007} in their classification of secondary stability modes. A collective behaviour can be observed in both modes: for $\alpha_v=0$, approximately twelve sub-units are involved in large-wavelength oscillations; whereas  a staggered pattern, characteristic of subharmonic instabilities, is observed for $\alpha_v=0.087$.


The cross-flow dynamics, displayed in figure \ref{fig:CF} on a group of three sub-units, is less sensible to collective behaviour, likely due to the smaller magnitude of the components. The streamwise-independent mode displays counter-rotating vortices. The left vortex seems to be constituted of the merger of two smaller co-rotating vortices. For $\alpha_v=0.087$, the vortical structures are sensibly different. The vortices seem to arise from the roll-up of a shear layer. This might have been anticipated, since it has been shown that secondary instability of swept flows can be related to Kelvin-Helmholtz instabilities \citep{wassermann_mechanisms_2002}. Furthermore, these secondary vortices are located on the edge of the boundary layer, in a region associated with strong wall-normal and spanwise shear layers. Ultimately, it appears that two distinct collective mechanisms are at play. The first is linked to streak instabilities, which push high-velocity fluid downwards near the walls while low-velocity fluid from the near-wall region is ejected higher in the boundary layer. The second instability mechanism is Kelvin-Helmholtz related, and generates secondary vortices through the roll-up of the shear layer located at the edge of the boundary layer.

\vspace{-0.1cm}
\section{Conclusions}\label{sec:CONCLUSIONS}

A new formulation for secondary stability analysis of linearly unstable flows has been proposed, which allows retrieving \textit{collective} amplification mechanisms involving several units of the primary instability. Such a method is based on the combination of a local two-dimensional stability analysis and  a Bloch waves formalism originally described in \citet{schmid_periodic_2017}. Within this framework, the linear stability of systems having a high number of degrees of freedom resulting from the repetition of several units of the primary instability, is investigated. Collective secondary modes, representing resonances or more complex multi-units interactions, are retrieved. 

The secondary stability of a swept boundary layer developing steady cross-flow vortices has been investigated within this framework. 
This flow case has been chosen since it allows a direct comparison with the Floquet analysis carried out by the work of \citet{fischer_primary_1991}. 

The instability dynamics seems to result from the competition between streamwise independent modes and detuned modes with detuning factor $\epsilon=0.35$. In the case of streamwise wavenumbers $\alpha_v > 0.087$, the secondary instability analysis can be accurately described by a low number of modes and can thus be retrieved by  Floquet analysis. Whereas, for smaller wavenumbers, a multi-modal instability arises. In this context, Floquet analysis fails to produce accurate stability results. The full spectra are retrieved, and unstable branches can be identified. The most unstable modes display collective behaviour, characteristic of multi-modal instabilities and observed in the DNS of \citet{hogberg_secondary_1998}. Notably, for streamwise independent perturbations, large wavelengths oscillations can be seen, while a staggered pattern, characteristic of a subharmonic transition, appears in the disturbance for $\alpha_v = 0.087$. The wavelength of the staggered pattern is imposed by the detuning factor.
Physically, two instability mechanisms have been identified: one linked to streaks and the circulation of high-momentum fluid towards the edge of the boundary layer. The other one appears linked to the roll-up of a shear layer and seems related to Kelvin-Helmholtz instabilities.

Ultimately, this approach appears of great benefit for secondary instability as it allows an easy identification of multi-modal instabilities with large wavelengths which could be difficult to retrieve through a Floquet analysis. Still, in the case of cross-flow vortices, more qualitative results could be obtained by using a nonlinearly saturated base flow. In general, studies on the secondary instability of fundamental coherent structures such as streaks could be completed and extended through the use of this framework. 

\vspace{.2cm}
\noindent \emph{Declaration of Interests. The authors report no conflict of interest.}




\bibliography{main}
\end{document}